\pdfoutput=1
\documentclass[%
 aip,
 pof,
 amsmath,amssymb,
 reprint,%
numerical,%
]{revtex4-1}

\usepackage{graphicx}
\usepackage{dcolumn}% Align table columns on decimal point
\usepackage{bm}% bold math
\usepackage{setspace}
\usepackage[left=1.5in, right=1.5in]{geometry}

\usepackage{tikz}
\usepackage{subfig}
\usepackage{amsthm}
\usepackage{algcompatible}
\usepackage{algpseudocode}
\usepackage{newfloat}

\DeclareFloatingEnvironment[
    fileext=loa,
    listname=List of Algorithms,
    name=ALGORITHM,
    placement=tbhp,
]{algorithm}

\newcommand{\defi}{:=}

\newcommand{\ctrans}{^T}  % complex conjugate transpose  (since we only deal with REAL data in fluids, set this to same as regular transpose for this paper)
\renewcommand{\Re}{\mathbb{R}}  % space of real numbers
\newcommand{\dmdK}{K} % DMD matrix
\newcommand{\dmdKt}{\tilde{K}} % Qx^T Qy A pinv(G)
\newcommand{\Qa}{Q_X} % Orthonormal basis for image of X
\newcommand{\Qb}{Q_Y} % Orthonormal basis for image of Y
\newcommand{\Gx}{G_X} % Gramian matrix for X, RxRx*
\newcommand{\Gy}{G_Y} % Gramian matrix for Y, RyRy*
\newcommand{\pinv}{^+}  % pseudo-inverse
\newcommand{\ord}{\mathcal{O}} % ``order of'' symbol, as in x~O( )
\newcommand{\ro}{r_o} % reset rank value in compressed DMD algorithm
\newcommand{\ra}{r_X} 
\newcommand{\rb}{r_Y} 
\newcommand{\ea}{e_X} 
\newcommand{\eb}{e_Y} 

\newcommand{\Rey}{\mathrm{Re}} % Reynolds number

\graphicspath{ {./figures/} }

\def\boxmatrix(#1,#2)#3{\tikz[baseline=(a.base)] \draw (0,0) node[minimum width=#2cm,
  minimum height=#1cm,draw,fill=black!20] (a) {$#3$};}

\begin{document}

%\begin{spacing}{0.75}
%\footnotesize{
%\preprint{Submitted to Physics of Fluids (Letters)}%AIP/123-QED}

\title[]{Dynamic Mode Decomposition for Large and Streaming Datasets}% Force line breaks with \\
%\thanks{Footnote to title of article.}

\author{Maziar S. Hemati}
 \email{mhemati@princeton.edu}
  \affiliation{Mechanical and Aerospace Engineering Department, Princeton University, NJ 08544, USA.}%Lines break automatically or can be forced with \\
\author{Matthew O. Williams}%
 \email{mow2@princeton.edu.}
\affiliation{ 
Program in Applied and Computational Mathematics, Princeton University, NJ 08544, USA.%\\This line break forced with \textbackslash\textbackslash
}%

\author{Clarence W. Rowley}
\email{cwrowley@princeton.edu}
\affiliation{Mechanical and Aerospace Engineering Department, Princeton University, NJ 08544, USA.}
%\\This line break forced% with 
%
%\altaffiliation{Program in Applied and Computational Mathematics, Princeton University, NJ 08544, USA.}
\date{\today}% It is always \today, today,
\pdfoutput=1
\begin{abstract}
%An article usually includes an abstract, a concise summary of the work
%covered at length in the main body of the article. It is used for
%secondary publications and for information retrieval purposes. 
%
We formulate a low-storage method for performing dynamic mode decomposition that can be  updated inexpensively  as new data become available;
this formulation allows dynamical information to be extracted from large datasets and data streams.
We present two algorithms: the first is mathematically equivalent to a standard ``batch-processed'' formulation; the second introduces a compression step that maintains computational efficiency, while enhancing the ability to isolate pertinent dynamical information from noisy measurements.
Both algorithms reliably capture dominant fluid dynamic behaviors, as demonstrated on cylinder wake data collected from both direct numerical simulations and particle image velocimetry experiments.
%
%\\
%\\
%Abstract must be $\le$100 words in length for \emph{Letters} submissions.
%
%\\
%Valid PACS numbers may be entered using the \verb+\pacs{#1}+ command.

\end{abstract}

%\begin{abstract}
%Abstract must be $\le$100 words in length for \emph{Letters} submissions.
%
%Valid PACS numbers may be entered using the \verb+\pacs{#1}+ command.
%\end{abstract}


\maketitle
\pdfoutput=1
Dynamic mode decomposition (DMD) is a data-driven computational technique capable of extracting dynamical information from flowfields measured in physical experiments or generated by direct numerical simulations.\cite{schmidJFM2010}
Since its introduction in 2008,\cite{schmidAPS2008} DMD has been used in the analysis of numerous fluid mechanical systems (e.g., bluff body flows, \cite{bagheriJFM2013} jet flows,\cite{rowleyJFM2009,jovanovicPF2014} and viscoelastic fluid flows\cite{grilliPRL2013}) and has gained increasing popularity owing to its ability to reveal and quantify the dynamics of a flow, even when those dynamics are nonlinear.\cite{rowleyJFM2009,mezicARFM2013}

DMD operates on snapshots of the flowfield (e.g., velocity, vorticity, pressure) and their time-shifted counterparts---obtained either from experiments or numerical simulations---to compute the eigenvalues (``DMD eigenvalues'') and eigenvectors (``DMD modes'') of a linear operator that best fits the associated dynamics in a least-squares sense.
The DMD modes represent spatial fields that often highlight coherent structures in the flow, while the associated DMD eigenvalues dictate the decay/growth rates and oscillation frequencies of these modes.
As such, access to DMD modes and eigenvalues enables a reconstruction of the dynamics associated with a given flowfield.
Other modal decomposition techniques, such as the commonly employed proper orthogonal decomposition (POD), only compute spatial modes associated with the flow.\cite{holmes2012}
Although spatial modes can offer valuable information regarding coherent structures and other flow qualities (e.g., in the case of POD, they determine the most energetic modes), characterizing the underlying dynamics relies upon projecting these spatial modes onto an assumed dynamical form.
DMD offers an advantage over these other modal decomposition techniques in that it computes \emph{both} spatial modes and their associated temporal behaviors, thus removing any guesswork associated with realizing a dynamical representation of the system.

To date, researchers have viewed DMD as a post-processing tool;
that is, a method that requires the entire experimental or computational dataset to be available prior to commencing analysis.
There are, however, circumstances in which an online and incrementally updatable algorithm for DMD would be advantageous over current batch-processing approaches.
Such a capability would allow DMD to be applied to streams of data, a paradigm shift that can be taken advantage of in numerous contexts, such as online flow analysis in conjunction with real-time particle image velocimetry (PIV).\cite{yuJACIC2006}
Moreover, a streaming DMD algorithm could be exploited for low-storage DMD analyses as well, since it would provide a means of performing DMD on large datasets by successively processing individual snapshots, one by one, without subsequently needing to store them all in memory.

In the present letter, we formulate a general framework that enables DMD computations to be updated incrementally as new snapshots become available.
We will introduce two algorithms: (1) a direct algorithm for updating DMD computations incrementally, which can be shown to be mathematically equivalent to ``batch-processed'' DMD, and (2) an extension of the direct algorithm that utilizes a POD basis for compression, which is well-suited for practical scenarios in which the data are corrupted by noise.
We demonstrate both algorithms on a canonical problem of laminar flow past a cylinder: the direct algorithm is used on data generated via direct numerical simulation (at Reynolds number $\Rey=100$, based on cylinder diameter), and the version with POD compression is applied to experimental PIV data obtained from water channel experiments at $\Rey=413$.
In both instances, we verify that the methods compute  dominant spatial modes and their associated temporal dynamics consistent with batch-processed DMD, but do so by working with the data incrementally.

%%% Local Variables:
%%% mode: latex
%%% TeX-master: "streamingdmd"
%%% End:

\pdfoutput=1
\newcommand\n{2}
\newcommand\m{1.5}
\renewcommand\r{0.75}

In formulating a means of updating DMD computations incrementally  as new
snapshots become available, we begin with the usual definition of the DMD
operator.\cite{schmidJFM2010,tuJCD2014}  That is, given pairs of snapshots $x_i\in\Re^n$
and $y_i\in\Re^n$ of the system states, spaced a fixed time-interval apart and
stored in the snapshot matrices $X \defi [x_1, x_2, \ldots, x_m]\in\Re^{n\times m}$
and $Y \defi [y_1, y_2, \ldots, y_m]\in\Re^{n\times m}$, one first computes a matrix
$\Qa\in\Re^{n\times \ra}$ whose columns form an orthonormal basis for the image
of $X$ (which has dimension~$\ra$); the DMD operator is then given by
\begin{equation}
        \dmdK = \Qa\dmdKt\Qa\ctrans,
        \label{eq:sdmdform}
\end{equation}
where $\dmdKt$ is an $\ra\times\ra$ matrix defined by
\begin{equation}
  \label{eq:Ktilde_standard}
  \underbrace{\boxmatrix(\r,\r){\vphantom{K}\smash{\dmdKt}}}_{\ra\times\ra} \defi \underbrace{\boxmatrix(\r,\n){\vphantom{Q}\smash{\Qa\ctrans}}}_{\ra\times n}\;
  \underbrace{\boxmatrix(\n,\m){Y}}_{n\times m}\;
  \underbrace{\boxmatrix(\m,\n){\vphantom{X}\smash{X\pinv}}}_{m\times n}\;
  \underbrace{\boxmatrix(\n,\r){\vphantom{Q}\smash{\Qa}}}_{n\times\ra}\;,
\end{equation}
where $X\pinv$ denotes the Moore-Penrose pseudoinverse of~$X$.  The DMD
eigenvalues and modes are then eigenvalues and eigenvectors of~$\dmdK$, and
these may be computed from the eigenvalues and eigenvectors of the much smaller
matrix~$\dmdKt$.  Note that, as the number~$m$ of snapshot pairs grows, the number of
columns of $Y$ (and rows of $X\pinv$) increases, so large numbers of snapshots
require large amounts of storage in order to compute $\dmdKt$.

In this letter, we are interested in situations in which we have access to only a single pair of snapshots $(x_i,y_i)$ at any given time, either due to computer memory limitations in storing large numbers of snapshots, or based on implementations on real-time data streams for which future snapshots are not yet available.
Our main contribution is to provide an alternative way of computing~$\dmdKt$,
such that it can be updated incrementally as new snapshots become available,
without storing previous snapshots.  To do this, we first determine orthonormal
bases for the images of $X$ and~$Y$, and stack these as columns of matrices $\Qa\in\Re^{n\times\ra}$ and~$\Qb\in\Re^{n\times\rb}$ (where $\ra$ and~$\rb$ denote the respective ranks of $X$ and~$Y$).  We then project the data vectors onto these coordinates, writing $\tilde X\defi\Qa\ctrans X$ and $\tilde Y\defi\Qb\ctrans Y$, and define new matrices $A\defi\tilde Y\tilde X\ctrans\in\Re^{\rb\times\ra}$ and $G_X\defi\tilde X\tilde X^T\in\Re^{\ra\times\ra}$.  Then using the identity $X^+=X\ctrans(XX\ctrans)\pinv$, the matrix $\dmdKt$ from~\eqref{eq:Ktilde_standard} may be rewritten as
\begin{equation}
  \underbrace{\boxmatrix(\r,\r){\vphantom{K}\smash{\dmdKt}}}_{\ra\times\ra} =
  \underbrace{\boxmatrix(\r,\n){\vphantom{Q}\smash{\Qa\ctrans}}}_{\ra\times n}\;
  \underbrace{\boxmatrix(\n,\r){\Qb}}_{n\times\rb}\;
  \underbrace{\boxmatrix(\r,\r){A}}_{\rb\times\ra}\;
  \underbrace{\boxmatrix(\r,\r){\vphantom{G}\smash{\Gx\pinv}}}_{\ra\times\ra}\;.
\label{eq:Ktilde}
\end{equation}
There are two main advantages of the formulation~\eqref{eq:Ktilde} over the
standard formulation~\eqref{eq:Ktilde_standard}: first, much less storage is
required, in the typical case that $\ra,\rb\ll m$; second, the required matrices
may be updated incrementally as new snapshots become available, as we describe
below.
%
%
%%In principle, $\Gx$ is symmetric and positive definite, and $\Gx\pinv=\Gx\inv$ can be computed from its Cholesky decomposition; 
%
%%however, in practice, $\Gx$ can be ill-conditioned, so $\Gx\pinv$ should be computed from the singular value decomposition (SVD) for the purposes of numerical stability.
%
%We note that  $\Gx\pinv$ can be computed from a Cholesky decomposition when it is numerically well-conditioned, since $\Gx\pinv=\Gx\inv$
%Because $\Gx$ is square, $\Gx\pinv=\Gx\inv$ when $\Gx$ is well-conditioned.  
%As a result, could be computed using a Cholesky decomposition.  
%However, when this is not the case, $\Gx\pinv$ is computed via the SVD for purposes of numerical stability.

Based on the definition in \eqref{eq:Ktilde}, we formulate a method to update $\Qa$, $\Qb$, and  $\dmdKt$ (and thus $\dmdK$) with the introduction of every new snapshot pair.
The first task is to determine whether the bases contained in $\Qa$ and $\Qb$ should be expanded. 
To do so, the residuals $\ea = x_i - (\Qa\Qa\ctrans)x_i$ and $\eb = y_i - (\Qb\Qb\ctrans) y_i$ are computed, and  
if $\|\ea\|$ or $\|\eb\|$ is greater than some pre-specified tolerance, then we expand $\Qa$ by appending $\ea/\|\ea\|$ to the last column of the matrix, and if needed, use an equivalent procedure to expand $\Qb$.  
The resulting orthonormal bases are identical to the ones that would be produced using the Gram-Schmidt process if $X$ and $Y$ were available in their entirety.
Next, we compute $\tilde x_i = \Qa\ctrans x_i$ and $\tilde y_i = \Qb\ctrans y_i$, which are the low-dimensional equivalent of the large snapshot pair.
The matrices that comprise $\dmdKt$ are then defined as $A = \sum_{j=1}^i \tilde y_j\tilde x_j\ctrans$ and $\Gx = \sum_{j=1}^i \tilde x_j \tilde x_j\ctrans$, which contain sums of outer products between the mode amplitudes from all previous observations. 
To update $A$ and $\Gx$ given a new snapshot pair $(x_i, y_i)$, we set $A \leftarrow A + \tilde y_i\tilde x_i\ctrans$ and $\Gx \leftarrow \Gx + \tilde x_i \tilde x_i\ctrans$, which incorporates  the last pair of outer products associated with the $i$-th snapshot pair.
We note that in order to account for the new basis element that was not present in the previous iterates, both $A$ and $\Gx$ must be ``padded'' with zeros whenever the size of $\Qa$ or $\Qb$ increases.
If $r= \max(\ra,\rb)$, the computational cost of each iterate is dominated by the orthogonalization step with a computational cost of $\ord(nr)$ if the DMD modes and eigenvalues are not required, and a cost of $\ord(nr^{2})$ if they are.
Therefore, this algorithm is particularly effective when $n$ and $m$ are large, but the ranks of $X$ and $Y$ are small.
%

\begin{figure}
\begin{center}
\subfloat[DMD Mode $(\lambda=0.998+0.0531i)$]{\includegraphics[width=0.5\textwidth]{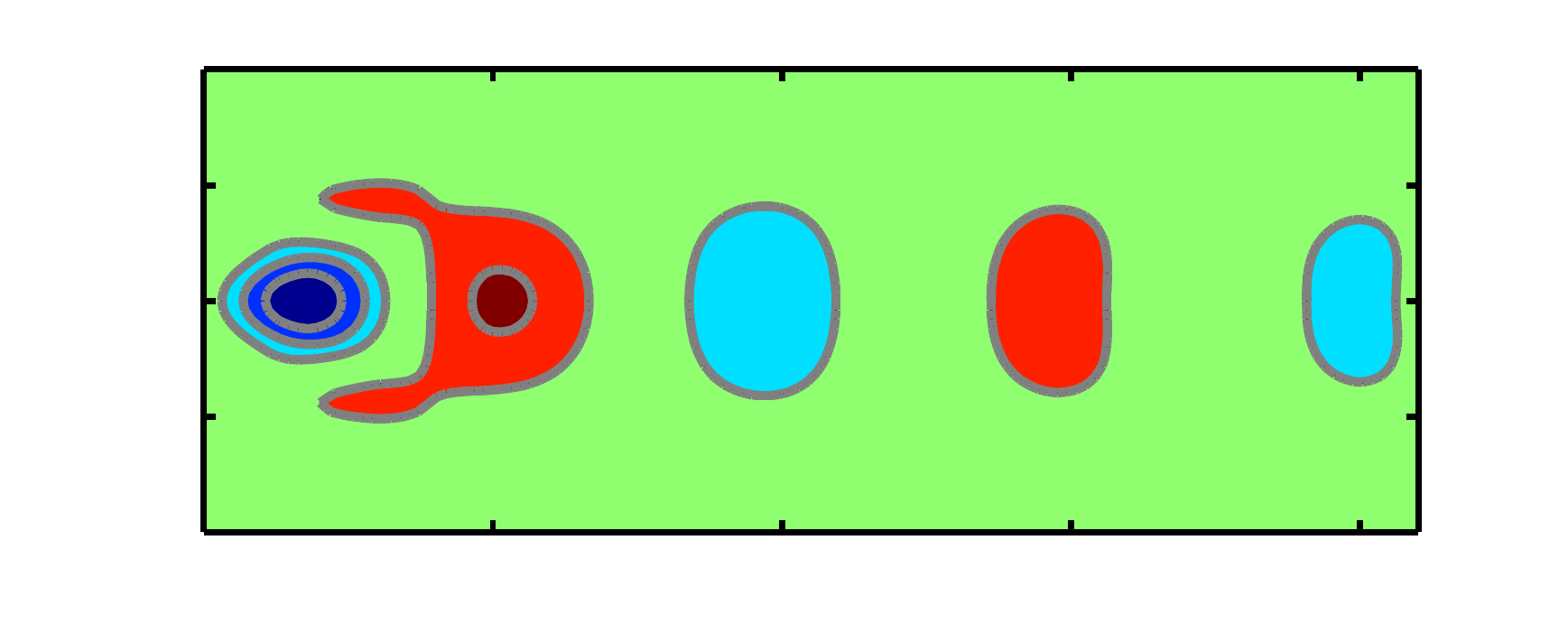}}
\subfloat[DMD Mode $(\lambda=0.994+0.106i)$]{\includegraphics[width=0.5\textwidth]{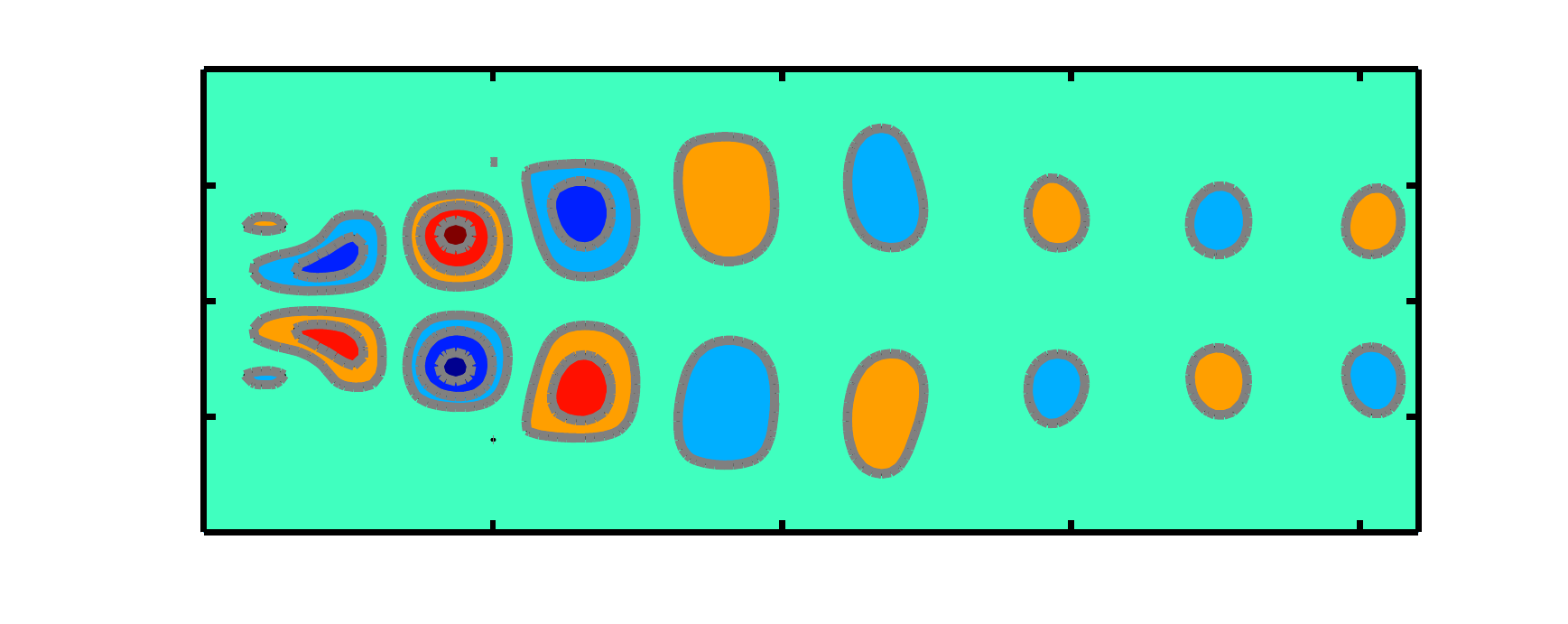}}
\end{center}
\caption{Incrementally computed and batch-processed DMD modes are identical.  Here, we plot the real components of the first two dominant oscillatory DMD modes corresponding to the incrementally updated  computations (plotted as gray contours of modal level sets) and the batch-processed results (plotted as filled contours between modal level sets).}
\label{fig:re100modes}
\end{figure}

%</%%%%%%%%%%%%%%%%%%%%%%%%%
% Direct Streaming Results
%%%%%%%%%%%%%%%%%%%%%%%%%%%
Now we demonstrate this incrementally updated DMD computation procedure and compare with results from a batch-processed approach by working with direct numerical fluids simulation data associated with two-dimensional laminar flow past a cylinder ($\Rey=100$ based on cylinder diameter).
We find that the DMD modes resulting from the incremental algorithm match those computed from a standard DMD implementation: Figure \ref{fig:re100modes} presents the first two dominant DMD modes, with the incrementally computed modes overlayed on top of the batch-processed modes.
Numerical considerations aside, even the less-dominant DMD eigenvalues and modes (not reported here) are also in close agreement.
%%%%%%%%%%%%%%%%%%%%%%%%%%/>

Although this direct algorithm is beneficial, unlike the demonstration on numerically generated data above, the snapshot data are often corrupted by noise in other practical settings;
that is, in many cases, $X$ and $Y$ can be decomposed into a low-rank component that contains the signal and a high-rank component that contains the noise, which results in $\ra,\rb\sim n$ when $m>n$.
This is problematic because the performance of the direct updating procedure is heavily dependent on the rank of the data. 
As such, a modification of the algorithm that allows the basis $\Qa$ and $\Qb$ to be {\em compressed} is now presented.
In this modified updating scheme, we make use of the same four matrices as in the direct updating procedure---$A$, $\Gx$, $\Qa$, and $\Qb$---and introduce a new matrix, $\Gy\defi\sum_{j=1}^i \tilde y_j\tilde y_j\ctrans$, to enable incremental POD compressions of the snapshots comprising $Y$.
%In this modified updating scheme, we make use of the same four matrices as in the direct updating %procedure---$A$, $\Gx$, $\Qa$, and $\Qb$---and introduce a new matrix, $\Gy$, which at the $i$-th data %point is $\Gy = \sum_{j=1}^i \tilde y_j\tilde y_j\ctrans$. 
%
As before, $\Gy$ can be updated easily from the previous iterate because $\Gy \leftarrow \Gy + \tilde y_i\tilde y_i\ctrans$. 
The $\Gx$ and $\Gy$ matrices are important for compression because $XX\ctrans = \Qa\Gx\Qa\ctrans$ and $YY\ctrans = \Qb\Gy\Qb\ctrans$, which are the matrices whose eigenvectors and eigenvalues give the POD modes and mode energies of $X$ and $Y$, respectively.\cite{holmes2012}
Furthermore, if $v_i$ is the $i$-th eigenvector of $\Gx$, then $\Qa v_i$  is the $i$-th POD mode of $X$, which eliminates the need to form either $XX\ctrans$ or $YY\ctrans$ explicitly. 
As a result, if the rank of either $\Gx$ or $\Gy$ exceeds some pre-specified value, then we modify $\Gx$, $\Gy$, $A$, $\Qa$ and $\Qb$ using the leading eigenvectors of $\Gx$ and $\Gy$.
Specifically, if $V_X$ and $V_Y$ have columns containing the leading eigenvectors of $\Gx$ and $\Gy$, then $\Gx\leftarrow V_X\ctrans \Gx V_X$, $\Gy \leftarrow V_Y\ctrans \Gy V_Y$, $A \leftarrow V_Y\ctrans A V_X$, $\Qa \leftarrow \Qa V_X$, and $\Qb \leftarrow \Qb V_Y$, which are the equivalent matrices as before, but now represented in a POD basis.

If $X$ and $Y$ have many small singular values, which is often the case when $X$ and $Y$ are generated by a low-rank process with a small noise component, then this truncation step can greatly reduce the dimensionality of the system with a minimal loss of accuracy; such a truncation is also critical to preserving the low-storage nature of our algorithm when $X$ and $Y$ are no longer low-rank, on account of any noise.
Because of the matrix multiplications needed to update $\Qa$ and $\Qb$, the computational cost of this step is $\ord(n\ro^2)$, where $\ro$ is a pre-specified maximum allowable matrix rank at which the truncation step occurs.
Due to the sequence of projections onto different POD bases, this algorithm is no longer equivalent to the standard DMD algorithm;
however, as we will demonstrate, this method produces dynamically relevant results, which are comparable to those computed from DMD directly.

A single iteration of the algorithm can be summarized as follows:
\begin{enumerate}
        \item For each new pair of data points $x_i$ and $y_i$, compute the residuals $\ea = (I - \Qa\Qa\ctrans) x_i$ and $\eb = (I - \Qb\Qb\ctrans) y_i$.
\item If $\|\ea\|>\epsilon$ or $\|\eb\|>\epsilon$, increase the dimension of the corresponding basis, $\Qa$ or $\Qb$, by appending an additional column $\ea/\|\ea\|$ or $\eb/\|\eb\|$, respectively, while zero-padding $\Gx$, $\Gy$, and $A$ to maintain dimensional consistency.
%        \item If $\|\ea\|>\epsilon$ or $\|\eb\|>\epsilon$, increase the dimension of the basis in $\Qa$ and $\Qb$.
        \item If either basis, $\Qa$ or $\Qb$, becomes too large (i.e., $\ra, \rb > \ro$), compute the leading eigenvectors of $\Gx$ and $\Gy$ (i.e., $V_X$ and $V_Y$, respectively), then set $\Gx\leftarrow V_X\ctrans \Gx V_X$, $\Gy \leftarrow V_Y\ctrans \Gy V_Y$, $A \leftarrow V_Y\ctrans A V_X$, $\Qa \leftarrow \Qa V_X$, and $\Qb \leftarrow \Qb V_Y$.
%        \item If either basis, $\Qa$ or $\Qb$, becomes too large (i.e., $\ra, \rb > \ro$), compress it, along with $A$, $\Gx$, and $\Gy$, by projecting onto the relevant set of POD modes that we construct from a subset of the eigenvalues and eigenvectors of $\Gx$ and $\Gy$.        
        \item Set $\tilde{x}_i = \Qa\ctrans x_i$ and $\tilde{y}_i = \Qb\ctrans y_i$, and let $\Gx \leftarrow \Gx + \tilde{x}_i \tilde{x}_i\ctrans$, $\Gy \leftarrow \Gy + \tilde{y}_i \tilde{y}_i\ctrans$, and $A \leftarrow A + \tilde{y}_i \tilde{x}_i\ctrans$.
        \item If the DMD modes and eigenvalues are required, compute the eigenvalues and eigenvectors of $A\Gx\pinv$.  If $v_j$ is the $j$-th eigenvector of $A\Gx\pinv$ then $\Qa v_j$ is the $j$-th DMD mode. 
\end{enumerate}
In total, if the DMD modes and eigenvalues are desired after every iterate, the computational cost of this algorithm is $\ord(n r^2)$ per iterate, where $r$ is on the order of the effective rank of $X$ and $Y$. 
In terms of storage, the algorithm requires matrices with $\ord(nr)$ entries; 
as a result, it will be computationally and memory efficient when $\ra, \rb \ll n$.  
More importantly, this is a ``single pass'' algorithm that does not require previous snapshots to be stored, thus making it useful for applications with large datasets or data streams for which $m\to\infty$.

%%% Local Variables:
%%% mode: latex
%%% TeX-master: "streamingdmd"
%%% End:

\pdfoutput=1
To highlight the benefits of POD compression for incrementally updated DMD computations in the face of noisy measurements, we apply the algorithm to the PIV data presented in Tu et al. (2014) for flow over a cylinder at $\Rey=413$.
The experiments, conducted in a water channel with precautions taken to minimize three-dimensional effects and surface wave interactions, sampled the velocity field at a rate of 20 Hz and yielded a final resolution of 135$\times$80 pixels.\cite{tuEIF2014}
A total of 8000 PIV snapshots were recorded with 8000 $\mu$s delay between exposures.

\begin{figure}[h!]
\begin{center}
\subfloat[Frequency Spectrum]{\includegraphics[width=\textwidth]{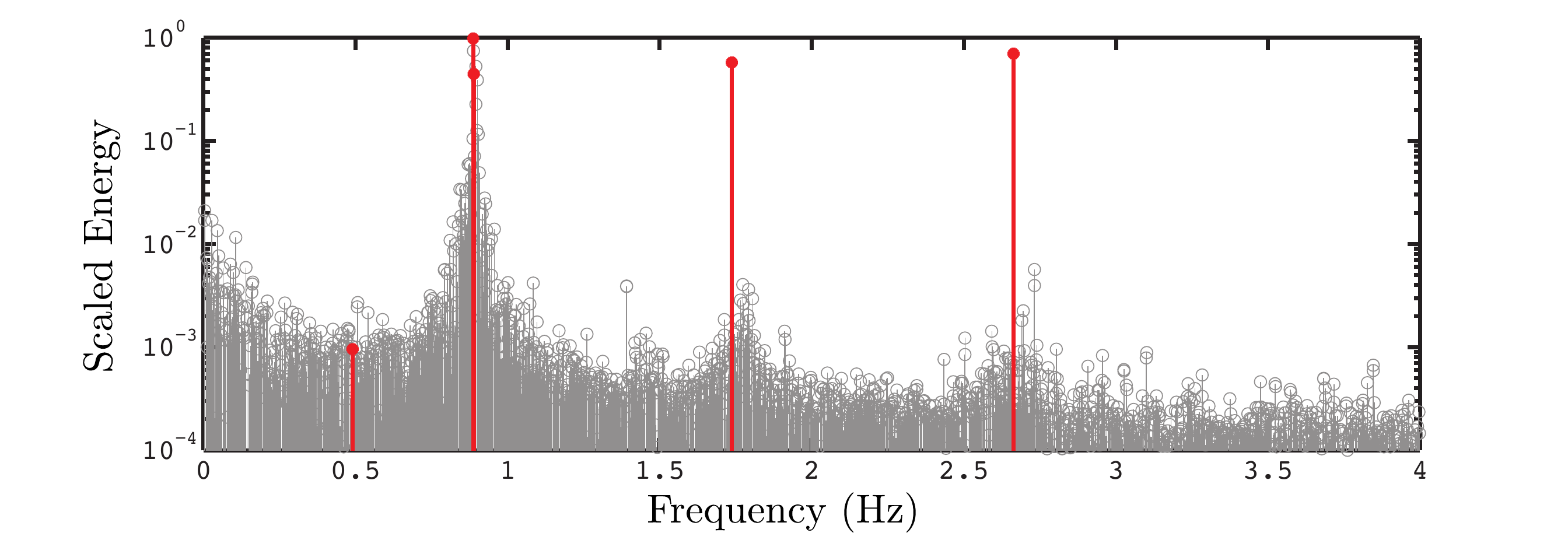}}\\
\subfloat[Batch-Processed \newline $f_1=0.888 \text{ Hz}$]{\includegraphics[width=0.33\textwidth]{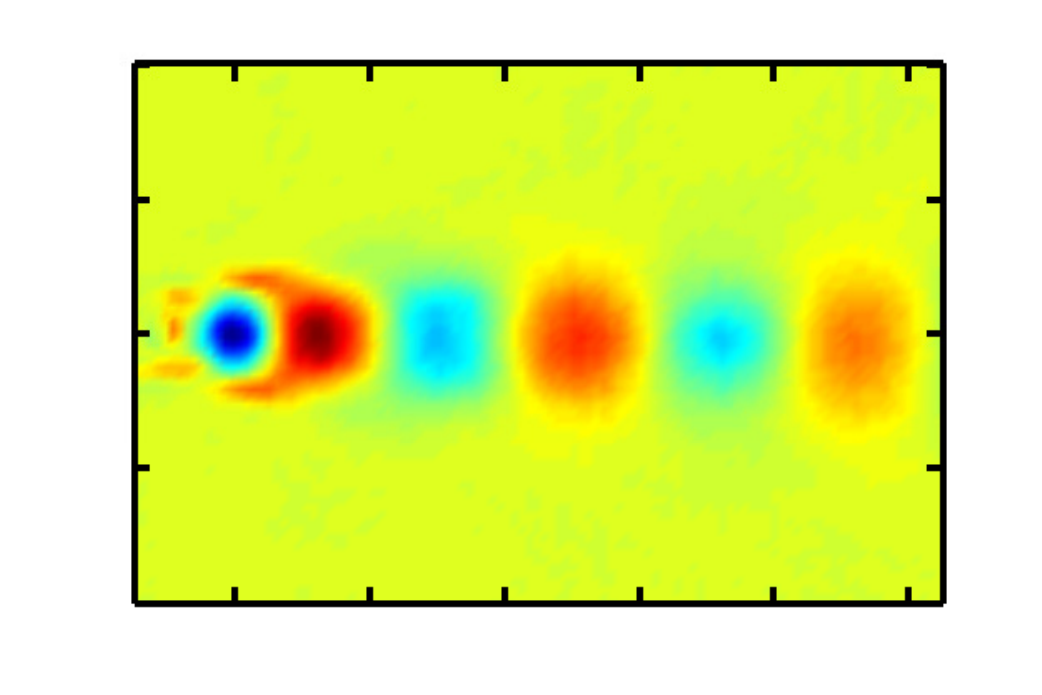}}
\subfloat[Batch-Processed \newline $f_1=1.774 \text{ Hz}$]{\includegraphics[width=0.33\textwidth]{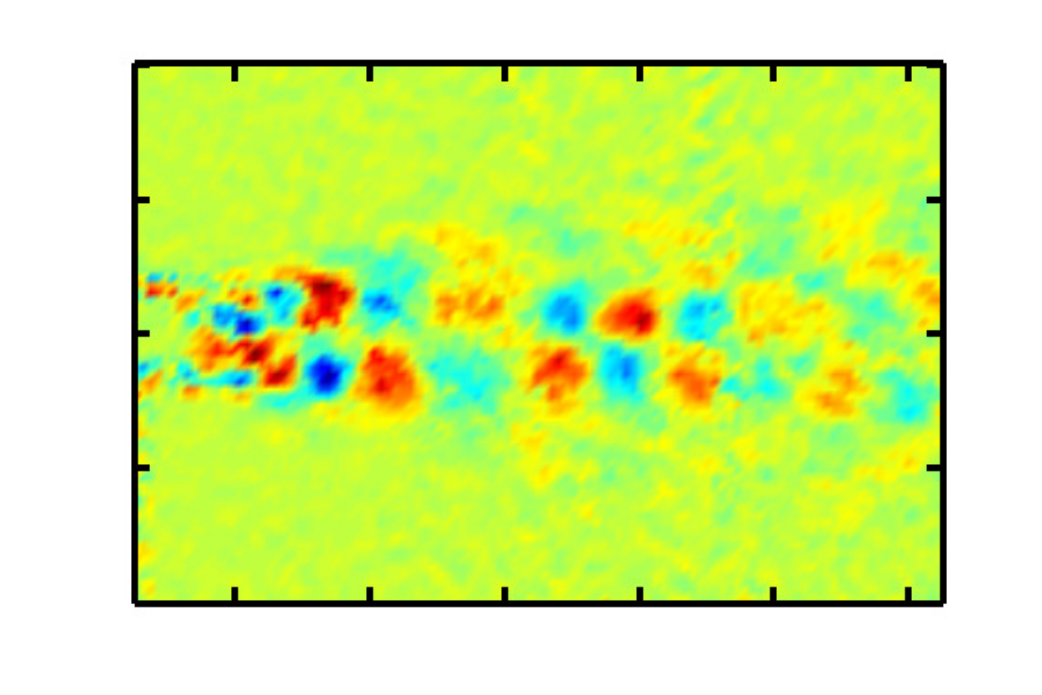}}
\subfloat[Batch-Processed \newline $f_1=2.732 \text{ Hz}$]{\includegraphics[width=0.33\textwidth]{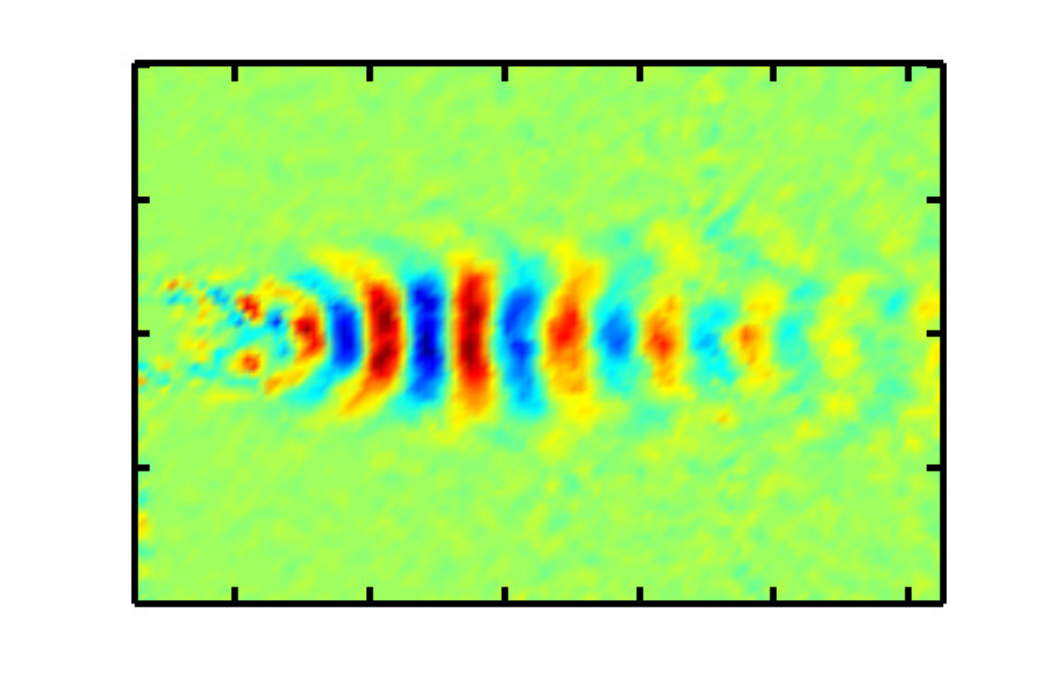}}\\
\subfloat[Incrementally Updated $f_1=0.887 \text{ Hz}$]{\includegraphics[width=0.33\textwidth]{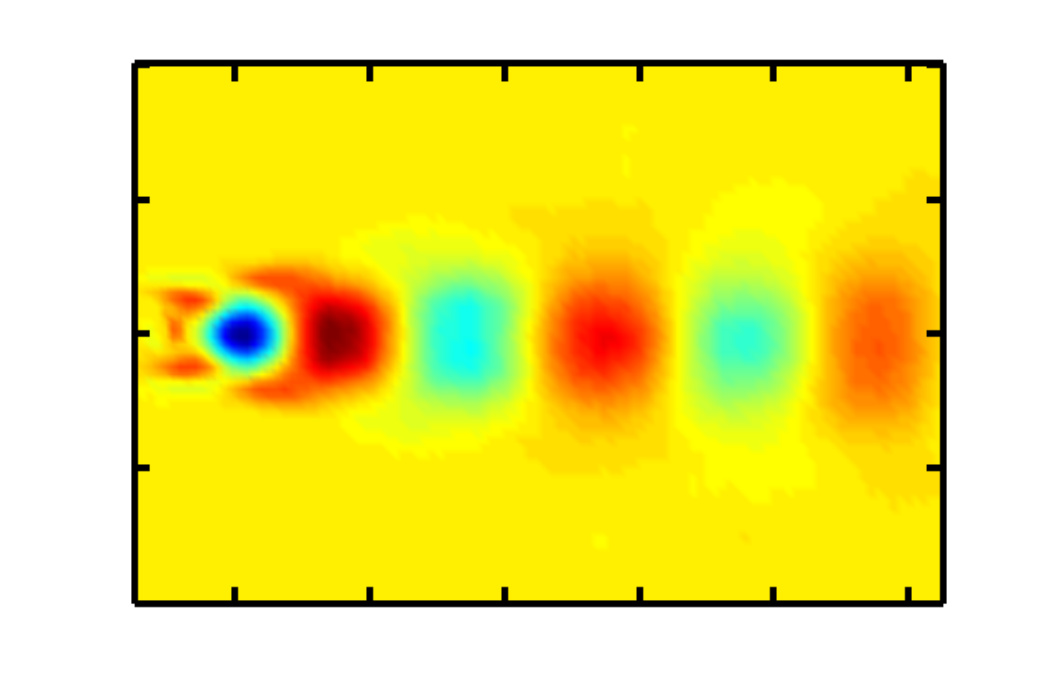}}
\subfloat[Incrementally Updated $f_2=1.737 \text{ Hz}$]{\includegraphics[width=0.33\textwidth]{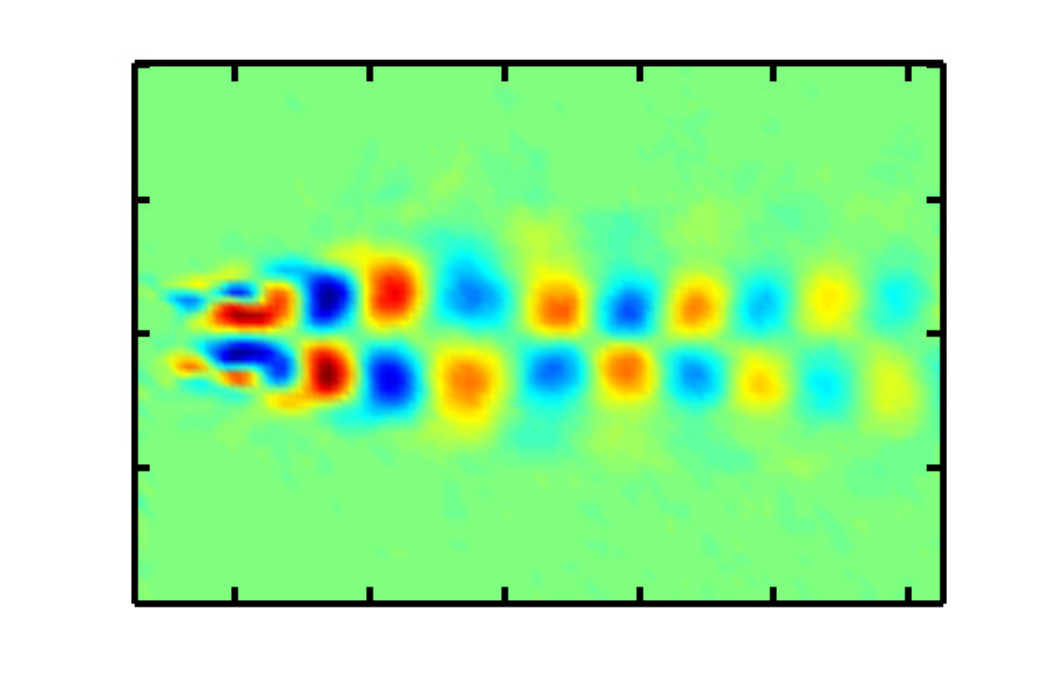}}
\subfloat[Incrementally Updated $f_3=2.664 \text{ Hz}$]{\includegraphics[width=0.33\textwidth]{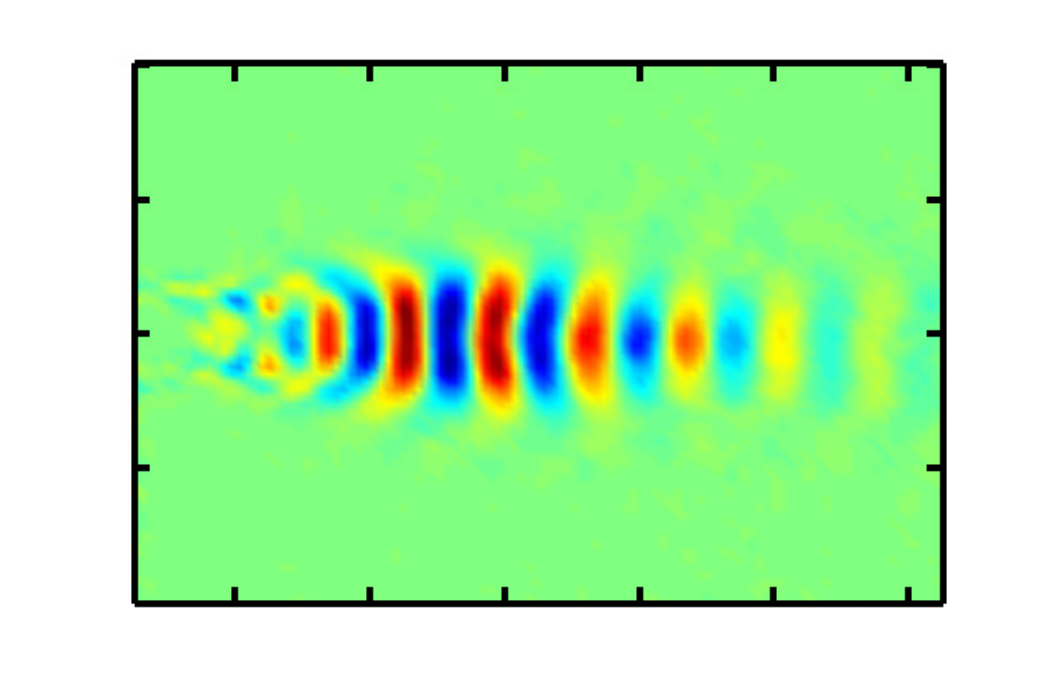}}
\end{center}
\caption{Updating DMD incrementally with POD compression yields approximately the same dominant frequencies and modes as batch-processed DMD.  In (a), we present the frequency spectrum of batch-processed DMD (hollow gray circles) with that corresponding to incrementally updated DMD with POD compression and $\ro=25$ (solid red circles). Tiles (b)--(d) present the real components of the dominant mode shapes computed from batch-processed DMD which can be compared with tiles (e)--(g) directly below, which present the real components of dominant modes computed via incremental updates and POD compression.}
\label{fig:freqspec}
\end{figure}

We applied our algorithm with POD compression to the PIV dataset on a personal computer and successfully identified the dominant DMD modes and their temporal characteristics.
In Figure \ref{fig:freqspec}, we overlay the incrementally computed frequency spectrum (with $\ro=25$) on top of the batch-processed DMD results of Tu et al. (2014), which required a parallel implementation of DMD on three computational cores to obtain.\cite{tuEIF2014,belsonACMTMS2014}
We note that although the POD compression step makes our algorithm ``different'' from DMD, it still yields relevant information about the dominant dynamics of a flow in an efficient manner.
Additionally, by comparing the results in Figure \ref{fig:freqspec}, it is clear that the updating procedure with POD compression succeeds in extracting smoother mode shapes than the batch-processed algorithm, since it is able to sift through and filter out the contributions from noise during the truncation stage.

%{\color{blue}{[POD DMD results on the full data stream] Present figures of modes and spectrum... also present a figure with rank versus number of snapshots processed to compare naive dmd and pod dmd (if space permits). }}
%
%
%{\color{blue}{Discussion of results...? Perhaps just a comparison with traditional DMD in terms of resulting modes/evals and computational time/memory requirements.
%
%Also discuss that POD DMD is not exactly the same as DMD/direct streaming DMD, but the results are still useful...}}

\pdfoutput=1
In this letter, we have presented two algorithms for performing DMD analysis in an incremental fashion as new data become available.
The first algorithm   approached this problem directly, with the assumption that the snapshot matrices were low-rank, and yielded DMD modes and eigenvalues that matched those computed from a post-processing implementation of DMD for the flow past a cylinder ($\Rey=100$) generated by direct numerical fluids simulations.
Indeed, it can be shown that the two algorithms are mathematically equivalent.
The second (more practical) algorithm relaxed the low-rank assumption imposed on the data matrices, instead relying upon a POD compression step to maintain a computationally efficient low-storage algorithm, even in the presence of noise.
The incrementally updated DMD algorithm with POD compression successfully extracted the dominant frequencies and associated modes for flow past a cylinder (Re=413) based on experimentally acquired PIV data.
Not only were the resulting mode shapes smoother than the batch-processed DMD calculations, but the incremental algorithm was implemented on a personal computer with little effort, while the batch-processed results required a parallel implementation with three computational cores.
The advantages of the incrementally updated DMD algorithm, both in terms of low-storage and potential for real-time implementation, will make DMD available in numerous contexts where it would not have been feasible previously.
For example, incremental updating will prove useful for online DMD analysis of real-time PIV or smoke/dye visualizations; it will also enable DMD analysis of massively large datasets that cannot completely reside in memory.
\\
\\
%%%%%%%%%%%%%%%%%%
%Acknowledgements
%%%%%%%%%%%%%%%%%%
We gratefully acknowledge Jessica Shang for providing access to the experimental PIV data for flow over a cylinder, as well as Scott T.M. Dawson and Jonathan H. Tu for sharing their insights on performing DMD analysis on such flows.
M.O.W. acknowledges support from NSF DMS-1204783.
C.W.R. and M.S.H. acknowledge support from the Air Force Office of Scientific Research.
\\
  \nocite{}
    \bibliography{streamingdmd}

%\input{./appendix.tex}
%}
%\end{spacing}

\end{document}